\documentclass{ws-procs9x6}

\begin{document}
\title{Perspectives on Issues Beyond the Standard Model\footnote{Opening
Plenary Talk, to appear in the Proceedings of SUSY 2003, held at the
University of Arizona, Tucson, AZ, 5-10 June 2003}}

\author{Gordon Kane}
\address{Michigan Center for Theoretical Physics\\
University of Michigan\\
Ann Arbor, MI 48109}

\maketitle

\abstracts{In this opening talk I first describe how we are entering a
data-rich era, and what clues we might soon have to physics beyond the
SM (on a time scale of several years). Then we turn to a number of the
basic issues we hope to explain, and discuss what level of theory may
address them, the SM or supersymmetric SM, or string theory, or
``other''. Next we consider some of the large number of fine-tunings we
seem to have, and how they may be clues, focusing on how the fine-tuning
of M$_{Z}$ and m$_{h}$ suggest the MSSM needs to be extended as the low
scale theory, and on how flavor physics may be a powerful probe of
string theory. Finally we examine using benchmark models to study all
these issues.}

\noindent We seem to be in a remarkable time for fundamental
physics. Perhaps finally the fundamental questions are now
scientifically formulated and are research problems. The parameters of
the universe have mainly been measured, and now need explaining. There
are good ideas about dark matter and dark energy and the matter
asymmetry and neutrinos. Quarks and leptons are almost certainly the
fundamental constituents -- they may be described as strings or
something else, but they are still quarks and leptons. Gauge theories
imply the forces. It is not that all the explanations exist, but
arguably we finally know what needs explaining.

There is a possible framework: M-theory $\Leftrightarrow$ string theory
$\Leftrightarrow$ 4D field theory + gauge groups + high scale
supersymmetry + quarks and leptons $\Leftrightarrow$ low scale effective
theory, the supersymmetric Standard Model and cosmology. There are still
gaps, and it could fail, but it could succeed.

We are entering a data rich era. Many experiments and facilities have
gotten underway recently or will soon after a decade or more of design
and arranging funding and construction. They include b-factories, cold
dark matter detectors, the upgraded Tevatron, $\nu $ properties
experiments, rare decays, $g_{\mu }-2,$ proton decay, WMAP, SDS, LHC,
etc.  We can expect to have ideas greatly focused and constrained by
data beyond the SM.

\bigskip

\noindent Having a framework to organize thinking is very important, but
it is unlikely (though it would be welcome) that major progress will
come from top-down study alone. What clues do we already have to physics
beyond the SM?

$\circ$ We have known of the matter asymmetry for a long time. There are
several approaches that seem to be able to explain how the universe can
evolve from an original matter-symmetric burst of energy to the observed
symmetry. All require physics beyond the Standard Models of particle
physics and cosmology as they are normally formulated. The problem is to
distinguish among them, which should be doable based on combinations of
theoretical and phenomenological analysis.

$\circ$ We have long known that non-baryonic dark matter exists, and
forms about 20\% of the universe. Such matter was predicted by
supersymmetry and by axions before it was settled from astronomy that
non-baryonic matter was needed. The SM cannot provide the dark matter.

$\circ$ Neutrino masses require physics beyond the SM as well, because
they require a new mass scale. If the smallness of neutrino masses is
from the favored see-saw mechanism, the new scale must be a high one and
the underlying theory must be a supersymmetric one to stabilize the
hierarchy.

These clues point in certain directions, though of course not uniquely
or clearly. Serious approaches should address them.

\bigskip

\noindent The clues we have just listed could all be explained by short
distance or cosmological phenomena. That is likely for neutrino masses,
and very unlikely for dark matter, but we cannot be sure. Some kinds of
data could guarantee that there is new physics at the electroweak
scale. The main possibilities for such a guarantee soon are the
following.

$\circ$ For large $\tan \beta$ the decay $B_{s}\rightarrow \mu \mu$
could have a branching ratio large enough to be observed at the Tevatron
even with its poor luminosity, and the LHC is better. In the SM it is
too small to be observed, so a signal is necessarily new physics, and
implies superpartners that are light enough to be produced at the
Tevatron and copiously at LHC.

$\circ$ The time-dependent CP asymmetry in $B\rightarrow \phi K_{s}$ is
predicted in the SM to equal $\sin 2\beta$ measured in $B\rightarrow
\psi K_{s}$ since the CPV in both arises from the initial B mixing.  At
present data from Belle suggests they are not equal. If that is
confirmed, again the deviation must be due to particles that are
observable at the Tevatron and LHC.

$\circ$ Another good window is the CP asymmetry in $B\rightarrow
s\gamma,$ which is quite small in the SM but need not be small when the
SM is extended.

$\circ$ It has long been known that g$_{\mu }-2$ is very sensitive to
supersymmetry breaking since it vanishes in the supersymmetry limit.
Both data and theory have improved since the first suggestions a few
years ago that a deviation from the SM may occur here. At present the
evidence leans toward a significant deviation, but some of the theory
arguments need to be confirmed. Such a deviation also could only arise
from virtual particles that could be directly studied at the Tevatron
and LHC.

$\circ$ The HEAT collaboration has reported an excess of cosmic ray
positrons with energies that could arise from WIMP annihilation, and are
consistent with neutralino LSP annihilation, particularly with higgsino
or wino type LSPs. These would have to be produced by non-thermal
mechanisms to give the relic density since they annihilate well and in
thermal equilibrium their number is too small to produce the observed
relic density. This is presently the only direct signal for dark matter.
Experimentally it is robust, having been observed in several balloon
flights of particle physics detectors, with systematics that changed
among the flights, but the background of cosmic ray positrons is
apparently not well enough understood to be confident this is a signal
of unexpected new physics.

$\circ$ A number of cold dark matter detectors are now taking data and
could report a signal soon.

$\circ$ Improved experiments for electric dipole moments (EDMs) are or
soon will be taking data. EDMs violate CP and are too small in the SM to
be observed, so any signal is physics beyond the SM.

$\circ$ The MiniBoone neutrino oscillation experiment at Fermilab should
report data in 2005 that settles whether there are three independent
neutrino mass differences. If three are indeed needed then sterile
neutrinos (without Standard Model gauge interactions) must exist, and
the implications for physics beyond the SM are profound. Many sterile
neutrinos exist in models, but very special circumstances are needed for
them to have typical neutrino masses and to mix significantly with the
normal neutrinos.

$\circ$ Proton decay is too small to observe in the SM, so a signal
would be an exciting proof of new physics, presumably a grand unified
theory.

$\circ$ In the SM lepton flavor violating decays such as $\mu
\rightarrow e\gamma$ or $\tau \rightarrow \mu \gamma$ are forbidden, but
they occur in all extensions of the SM. Tau decays are studied at $ B $
factories, with new levels of sensitivity as the luminosity
increases. Improved experiments for studying $\mu $ decays are underway.

$\circ$ The Tevatron can produce heavier particles than any previous
facility. Its luminosity has not reached the levels it could have, so
the event rate may be too small for signals to be seen over backgrounds,
but signals may occur in the next few years. LHC will be a factory for
new particle production, for superpartners and higgs bosons. Signals
will be observable, but interpretation may be difficult.

\bigskip

\noindent One should also keep in mind that what constitutes ``data''
can be rather subtle, and often not noticed until thinking is ready for
it. For Newton that the moon stays in orbit was crucial data, implying
there must be a force acting on it toward the center of the earth. The
flat old universe was important data that supported or suggested
inflation. Today several non-standard pieces of data exist and have
powerful implications that are sometimes ignored:

$\circ$ the hierarchy problem, that generally in a quantum theory the
Higgs boson mass will have Planck scale quantum corrections that imply
it will be raised to the Planck scale. Then all masses proportional to
it are also raised to that scale, the lepton and quark and W,Z masses.

$\circ$ LEP and other precision data found no significant deviations
from SM predictions at the 0.1-1\% level. In quantum theory whatever
effects allow the hierarchy to be maintained between the Planck scale
and the electroweak scale that characterizes the actual masses will
enter into all observables, so such effects must be weak and decoupled
for general reasons given the LEP and other precision data,

$\circ$ since a Higgs boson was not discovered the simple and elegant
interpretation of electroweak symmetry breaking and perhaps gauge
coupling unification is showing some fine-tuning that raises important
questions -- we will return to this below. Similar fine-tuning issues
arise from the absence of electric dipole moments for electrons and
quarks and from the absence of flavor changing neutral currents.

\bigskip

\noindent Another way to think about issues is to ask where various
important questions can be addressed. Note we are only asking where the
questions can be \textit{addressed,} rather than answered -- we know
lots about where questions can't be answered, but not so much about
where in the theory they will actually be answered. By addressed I mean
not just incorporated, but actually explained in terms of more basic
structures. For example, the SM can incorporate 3 families, and some CP
violation, but it does not explain the origin of these. Another way of
thinking of it is that addressing the question means that the associated
physics would have arisen or led to the phenomenon even if we did not
already know it was there. It's fun to make a table to help think about
where various issues are addressed.


\begin{table}[ph]
\tbl{}
{\footnotesize
\begin{tabular}{@{}crrrr@{}}
\hline
{}&{}&{}&{}&{}\\[-1.5ex]
QUESTION & SM & SSM & String Theory & Other\\[1ex]
\hline
{}&{}&{}&{}&{}\\[-1.5ex] 
what is matter & $\surd$ & {} &{} &{}\\[1ex] 
what is light & $\surd$ &{} &{} &{}\\[1ex] 
what interactions give our world & $\surd$ &{} &{} &{}\\[1ex] 
what stabilizes $m_{pl}/m_{W}$ &{} & $\surd$ & $\surd$\hspace{.5cm} & ? 
\hspace{.8cm}\\[1ex] 
gauge coupling unification &{} & $\surd$ &{} &{}\\[1ex] 
explain EWSB, the Higgs mechanism &{} & $\surd$ &{} & ? \hspace{.8cm}\\[1ex] 
how is supersymmetry broken &{} &{} & $\surd$\hspace{.5cm} &{}\\[1ex] 
is there a grand unified theory &{} &{} & $\surd$\hspace{.5cm} &{}\\[1ex] 
proton decay &{} & $\surd$ & $\surd$\hspace{.5cm} &{}\\[1ex] 
what is the origin of flavor physics &{} &{} & $\surd$\hspace{.5cm} &{}\\[1ex] 
values of q,l$^{\pm }$ masses &{} &{} & $\surd$\hspace{.5cm} &{}\\[1ex] 
values of neutrino masses &{} &{} & $\surd$\hspace{.5cm} & ? \hspace{.8cm}\\[1ex] 
physics of $\mu $ &{} &{} & $\surd$\hspace{.5cm} &{}\\[1ex] 
R-parity conservation &{} &{} & $\surd$\hspace{.5cm} &{}\\[1ex] 
cold dark matter &{} & $\surd$ &{} & $\surd$\hspace{.8cm}\\[1ex] 
value of tan$\beta $ &{} & $\surd$ & $\surd$\hspace{.5cm} &{}\\[1ex] 
weak CPV &{} &{} & $\surd$\hspace{.5cm} &{}\\[1ex] 
strong CPV &{} &{} & $\surd$\hspace{.5cm} & ? \hspace{.8cm}\\[1ex] 
baryogenesis &{} & $\surd$ &{} &{}\\[1ex] 
what is the inflaton &{} & $\surd$ & $\surd$\hspace{.5cm} &{}\\[1ex] 
cosmological constant is small &{} & $\surd$ & $\surd$\hspace{.5cm} &{}\\[1ex] 
what is the dark energy &{} & $\surd$ & $\surd$\hspace{.5cm} &{}\\[1ex] 
what are quarks and leptons &{} &{} & $\surd$\hspace{.5cm} &{}\\[1ex] 
what is electric charge &{} &{} & $\surd$\hspace{.5cm} &{}\\[1ex] 
how does space-time originate &{} &{} & $\surd$\hspace{.5cm} &{}\\[1ex] 
how does the universe originate &{} &{} & $\surd$\hspace{.5cm} &{}\\[1ex] 
why does quantum theory give the rules &{} &{} & $\surd$\hspace{.5cm} &{}\\[1ex]
\hline
\end{tabular}\label{table 1}}
\vspace*{-13pt}
\end{table} 

\bigskip

\noindent Here SSM is Supersymetric Standard Model. We don't know if
string theory addresses all the issues we assign to it, but we can be
hopeful. Often people say too little is known about string theory to
make progress in answering major questions about the real world like
these. We know that string theory and string theology exist, but do
string cosmology and string phenomenology really exist? Perhaps it is
better to approach it the opposite way -- if string theory is to be
relevant it is necessary to begin to do string cosmology and string
phenomenology, to try to work on these issues.

The column ``other'' is pretty empty. Large extra dimensions and other
approaches of recent years have not done well at actually addressing the
issues in the table. Perhaps they address the hierarchy problem, but no
more successfully than supersymmetry -- in one case one must assume the
large dimensions are of order the weak scale, and in the other that
superpartners are of order the weak scale. Once that assumption is made
the supersymmetric approach then predicts without additional assumptions
both gauge coupling unification and radiative electroweak gauge symmetry
breaking, while non-supersymmetric approaches explain nothing
additional.  Some can also deal with electroweak symmetry breaking, but
only with additional special assumptions such as boundary
conditions. Thus I will stay in the context of the supersymmetric
Standard Model and not discuss alternatives.

To deal with the question of whether too little is known about string theory
to make progress it helps to recall how little data and theory was needed to
formulate the Standard Model. Basically what was known experimentally when
the SM was successfully formulated was that there were quarks, 2 neutrinos,
V-A currents, parity violation (chiral fermions in modern language), weak
interactions were weak, the hadron spectrum, and early scaling in deep
inelastic scaling. One can argue that we have information of similar
quality today, as exemplified by the questions we can ask in the table.
Some theoretical structures were also known, gauge theories, the Higgs
mechanism, the renormalizability of the electroweak theory, and asymptotic
freedom. Many theoretical questions, particularly non-perturbative ones,
could not be answered, but did not prevent basically formulating the theory.
Perhaps it is optimistic to think we are in a similar situation today, but
it is at least defendable that the concepts and frameworks exist for clever
physicists to make major progress in addressing the issues of the table.

One area that should be emphasized is flavor physics. The SM and
supersymmetric SM nicely accommodate flavor physics. They do not require
or explain it. While string theory has not yet solved flavor physics
problems, it does address them in the sense we described above. String
theories in ten dimensions have properties that can lead to flavor
physics in the 4D world, to quarks and leptons and a few replicating
families, and to Yukawa matrices in the superpotential that may explain
the masses. If we only knew today that the world we see was made of
electrons and up and down quarks, string theory would force us to think
about whether other particles existed, in families, even though
understanding is not yet deep enough to call the existence of families a
crucial prediction. Again one can turn it around. Flavor physics may be
the area of particle physics most directly related to string theory, the
area where data may most directly point to the structure of string
theories. String theories have U(1) and perhaps other symmetries that
may be the symmetries needed to define flavor and to understand fermion
masses.

\bigskip

\noindent Another area that may be providing clues to point beyond the
MSSM to the actual low scale theory, or to special properties of the
MSSM, is several fine-tunings that are increasingly apparent as data
improves. By ``fine-tunings'' I mean phenomena that need
explaining. Lots of quantities have basically natural values, and others
do not. Given $M_{W},$ $M_{Z}$ has a natural value. It would be nice to
explain the value of $\theta _{W}$ that relates them, but there is no
sense that there is a fine-tuning here. If fermion Yukawa couplings all
had values either of order unity or a few per cent there would be no
sense of fine-tuning, since a theory that gave tree level Yukawas of
order unity or zero is reasonable, and occurs in some forms of string
theory, and higher order terms all of the same order is reasonable. But
the double hierarchy of fermion masses, with the heaviest member of the
families varying by two orders of magnitude, and masses in each family
varying by over an order of magnitude, requires explaining. Similarly,
if Higgs boson masses had been below about 100 GeV they would have been
natural, but if they are really over 110 GeV they need special
explaining.

\noindent It is worth listing fine-tunings since they can be seen as
clues to the underlying theory.

$\circ $ Why is the cosmological constant so small?

$\circ $ Why is $\Omega_{DE}\sim \Omega_{DM}\sim \Omega_{B}?$

$\circ $ Why is strong CP violation so small?

$\circ $ The Higgs hierarchy problem.

$\circ $ In supersymmetry, the $\mu $ problem.

$\circ $ M$_{Z}?$ In supersymmetry the natural scale for the Z mass is
that of the soft parameters, a few hundred GeV or more.

$\circ $ m$_{h}?$ The higgs mass is exponentially sensitive to soft
parameters in supersymmetry.

$\circ $ The natural value of tan$\beta $ is of order unity. There is
some evidence it may be large, which requires fine tuning.

$\circ $ In supersymmetry flavor changing neutral currents are naturally
much larger than observed ones.

$\circ $ Why is $m_{e}\ll m_{t}?$

$\circ $ If neutrino masses are hierarchical why are they so different?

Most of these are of course well known, and it is familiar to wonder what
they are telling us. One reason to encourage string phenomenology is that
all of them can be addressed from string theory, so it is easier to ask
there what sort of clues they are providing. The allowed region of MSSM
parameter space is actually quite small, and may be telling us a great deal
about the underlying theory.

\bigskip

\noindent I want to focus briefly on two of them, M$_{Z}$ and m$_{h}$
and the connection between them, because more recent data has
exacerbated the situation here. What do we know about the supersymmetry
soft breaking parameters? They have to be of order the TeV scale in
order to eliminate the hierarchy problem, but that is not very
precise. Both gauge coupling unification and radiative electroweak
symmetry breaking work qualitatively, and depend on the same soft mass
parameters plus $\mu ,$ once the hierarchy problem is solved. But we can
examine them more carefully. A way to think about them is to recognize
that the only relation we have that links the supersymmetry soft
breaking terms to a measured number comes from explaining the Z mass
with radiative electroweak symmetry breaking,

\[
M_{Z}^{2}\approx -2\bar{\mu}^{2}+6\bar{M}_{3}^{2}+...
\]

where the dominant terms on the RHS are shown, $M_{3}$ is the SU(3) soft
breaking mass, the bars above $\mu $ and $M_{3}$ mean they are evaluated
at the high scale, and the coefficients depend on various kinds of
assumptions but some such relation is robust. Existing data on chargino
production requires that $\bar{\mu}$ be larger than about M$_{Z},$ and
that $M_{3}$ be even larger. Thus this relation requires differences of
large numbers to ``explain'' $M_Z$. A number of ways out of this,
including lowering the high scale, looking for relations between
$\bar{\mu}$ and $\bar{M}_{3},$ extra matter, etc, do not change the
problem that the LHS seems to be small compared to the RHS.

This problem has been around for a long time, but it has become more
serious as ways out were examined and found not to work. It has become
worse recently as lower bounds on the Higgs mass have tightened, because
m$_{h}$ is also calculable in terms of soft parameters. The tree level
higgs boson mass is given by

\[
m_{h}^{2}\leq M_{Z}^{2}\cos ^{2}2\beta
\]

in the MSSM, so any of $m_{h}$ above M$_{Z}$ must come from radiative
corrections involving soft parameters. To get $m_{h}$ up to 115 GeV
(remember they add quadratically) requires large soft parameters, in
particular large $M_{3},$ which worsens the situation for M$_{Z}!$ There
is also great sensitivity here. For example, increasing $m_{h}$ from 112
to 115 GeV doubles the needed $M_{3}!$

What does this imply? One possibility is that the higgs mass is actually
lighter than 115 GeV but because the cross section is suppressed or the
signature is non-standard no signal was seen at LEP. This is possible,
but the kinds of models needed for it to happen are not so attractive
that any seem worth pursuing in the absence of confirming data. Another
possibility that seems well worth taking seriously is that the correct
\textit{low scale theory} is not the MSSM but some extended theory. Note
that only extensions that significantly change the higgs sector, so as
to modify both the calculation of M$_{Z}$ and of m$_{h}$, can be
relevant. Both are very sensitive to the smallness of the coefficient of
the fourth power of the higgs field in the higgs potential, so modifying
that is a way to focus. That supersymmetry fixed that coefficient in
terms of gauge couplings was a great success of supersymmetry, but the
result is $\sim (g_{1}^{2}+g_{2}^{2})/8 \ll 1$ rather than $ \sim 1$ as
suggested by the data. Of course one could say that the needed fine
tunings in supersymmetry suggest that supersymmetry is not the correct
approach, but all other approaches are so much more fine tuned in many
ways that we clearly want to improve on minimal supersymmetry rather
than reject it.

\bigskip

\noindent Another fine-tuning that is part of flavor physics is the
issue of the phases of the soft parameters. They are complex masses, and
introduce a number of new phases, which lead to a number of new CP
violation effects.  Most dramatic are electric dipole moments for
electrons and quarks, that are expected to be one to two orders of
magnitude larger than the current limits. One could interpret this to be
telling us that the relevant soft phases are real, in which case we
would have a significant clue about the underlying theory. The soft
phases depend on the superpotential and the Kahler potential directly
from the string theory, and on supersymmetry breaking, which could be
from F-term vevs. So far, however, no principle or argument has been
found that implies the phases should be small.  Phenomenologically, the
phases that enter so far could be small from effective cancellations
that appear natural in the high scale theory but seem fine-tuned in the
low scale theory (examples are known). If so non-zero EDMs would have to
appear as experiments improve a little more, and then the results would
point us in quite a different direction about the phase structure of the
high scale theory.

So how should we proceed? There is a great deal of interesting and
exciting work to do. First of course it is necessary to get as much
experimental and phenomenological information about the soft breaking
Lagrangian as possible, the soft masses and $\mu $ and the soft phases,
from colliders and b-factories and rare decays and EDM experiments and
dark matter experiments. This is difficult because the information from
all of these depends on a number of soft parameters at once, and we
never can have enough observables to invert the equations to measure a
single soft parameter, or tan$\beta ,$ until we have an electron linear
collider -- and that is at least 17 years off if one tracks through the
time line required by funding and site and construction issues. So
considerable clever thinking is needed by experimenters and theorists to
untangle the low scale theory. Further, even if we can learn some or
most of the low scale theory there are many obstacles to extrapolating
to the high scale effective Lagrangian. These include possible
intermediate scale matter, not knowing the several high scales that may
exist, not knowing the full gauge group, possible $D^{\prime}$ terms
that affect scalar masses, extra Higgs and neutralinos, and more. Again,
considerable challenging work is needed, here particularly by
phenomenological theorists, to learn to make progress.  Finally even
after we learn much of the high scale Lagrangian we have to recognize
its implications for connecting to an underlying string theory, what it
is telling us about where the world is in the M-theory amoeba.

\bigskip

\noindent A very good way to study all these issues is to study
``benchmark'' models.  Basically one can begin with a string theory and
start to work out what it predicts for the superpotential, the Kahler
potential, and the soft breaking Lagrangian that together determine the
properties of the observable particle physics world. Of course today
assumptions must be made to proceed at a number of stages, so one is
constructing a model. Doing so is very good for improving one's
understanding of the theory at many levels. Once a model is obtained one
can simulate what sort of discoveries of physics beyond the SM would be
made. Then one can pretend one only has those phenomena, with
experimental errors, and try to reconstruct the high scale theory one
knows one started from. If the model has intermediate scale matter one
can see how to learn that from the low scale data. Eventually a number
of clever people could learn how to proceed with real data as we learn
that. One can sometimes skip steps -- if one plots graphs of relations
among several low scale quantities one finds that various high scale
theories lead to different relations in characteristic ways, as an
example of how innovative studies could teach us techniques to overcome
obstacles. I have described using benchmark models to study collider
physics and some aspects of the soft breaking Lagrangian, and a similar
approach could be used to relate flavor phenomena and string theory. Of
course one could wait until there is data, but it is hard to recognize
when enough incomplete data has accumulated to make progress, while if
we carry out these studies now we may find progress earlier than we
expect.

\bigskip

\noindent {\bf Acknowledgments:} I would like to thank the organizers,
and particularly Keith Dienes, for a stimulating meeting and kind
hospitality. I appreciate very much a number of discussions and
collaborations, particularly with Lisa Everett, Steve King, Joe
Lykken. Pierre Binetruy, Brent Nelson, Ting Wang, and Liantao Wang, on
the topics here. Since this is an introductory talk for the meeting and
covers many topics, rather than a review, I have chosen to rigorously
not provide references not only to the field broadly since that would
have required hundreds of references, but also to specific results. I
apologize to those who would have been referenced, and particularly to
my collaborators. The reader can easily find the relevant references to
pursue each topic.

\end{document}